\pdfoutput=1

\documentclass{article}

 \usepackage{arxiv}

\usepackage[utf8]{inputenc} 
\usepackage[T1]{fontenc}    
\usepackage{hyperref}       
\usepackage{url}            
\usepackage{booktabs}       
\usepackage{amsfonts}       
\usepackage{nicefrac}       
\usepackage{microtype}      
\usepackage{lipsum}
\usepackage{graphicx}
\usepackage{amsmath}
\usepackage{subcaption}
\usepackage{caption}
\usepackage{placeins}

\usepackage{bm} 

\hypersetup{
    colorlinks = true,
    linkcolor = black,
    anchorcolor = black,
    citecolor = black,
    filecolor = black,
    urlcolor = blue
    }

\DeclareMathOperator*{\argminA}{arg\,min}

\title{End-to-End Amp Modeling:\\ From Data to Controllable Guitar Amplifier Models}

 \author{
  Lauri Juvela,
  Eero-Pekka Damskägg, 
  Aleksi Peussa,
  Jaakko Mäkinen,
  \And
  Thomas Sherson,
  Stylianos I. Mimilakis,
  Athanasios Gotsopoulos%
  \\
   Neural DSP Technologies\\
   Helsinki, Finland\\
 }
\newcommand{\blackboxamp}{Matchless DC-30 } 

\begin{document}
\maketitle
\begin{abstract}
This paper describes a data-driven approach to creating real-time neural network models of guitar amplifiers, recreating the amplifiers' sonic response to arbitrary inputs at the full range of controls present on the physical device. 
While the focus on the paper is on the data collection pipeline, we demonstrate the effectiveness of this conditioned black-box approach by training an LSTM model to the task, and comparing its performance to an offline white-box SPICE circuit simulation.
Our listening test results demonstrate that the neural amplifier modeling approach can match the subjective performance of a high-quality SPICE model, all while using an automated, non-intrusive data collection process, and an end-to-end trainable, real-time feasible neural network model.

\end{abstract}

\renewcommand{\runningtitle}{End-to-end Amp Modeling}
\section{Introduction}
\label{sec:intro}
Digital guitar amplifier modeling is the process of recreating the behaviour of analog amplifier circuitry in a computer program.
A successful digital model recreates the amplifier response to user input\footnote{By user input we refer to the digitized audio signal of the musical instrument performance of the user and to the device-specific controls that affect auditory characteristics.}
in a way that is perceptually indistinguishable from the analog reference, both in terms of sound and feel.
Therefore, modeling technologies enable a vast range of users, spanning from amateur to professional musicians, audio engineers, and producers, to conveniently and reliably have tonal captures of their desired devices. Another benefit of modeling technologies is that they enable a wide access to analog devices that are scarce and expensive to acquire, i.e., boutique and vintage devices.

In a physical amplifier circuit, complex interactions take place between reactive elements, such as capacitors and inductors, and non-linear elements, such as vacuum tubes, transistors, and diodes.
These non-linear interactions constitute a dynamic system and are largely responsible for the desired sonic characteristics of guitar amplifiers.
White-box models based on nonlinear circuit simulation aim to directly recreate this complex analog system in the digital domain \cite{pakarinen2009-review-of-tube-amp-modeling, giampiccolo2021multiphysics, najnudel2021power-balanced, giampiccolo2022parallel}.
The circuit is represented as a system of nonlinear 
differential equations
, which are solved using numerical methods.
Assuming knowledge of the modeled circuit, these methods can produce accurate emulations, particularly for solid-state circuits with relatively few components.
However, circuits with many reactive elements coupled with nonlinearities can become prohibitively expensive to simulate in real time.
In addition, the behavior of vacuum tubes and transformers is in most cases difficult to describe or approximate analytically \cite{zoelzer2011dafx}.

A more lightweight modeling approach is to approximate the behavior of guitar amplifier stages using a combination of filtering and waveshaping. 
In discrete time, such approximation can be realized as an ensemble of linear time-invariant (LTI) digital filters and static nonlinear functions.
Determining the transfer functions of the filtering operations and tuning the non-linear behavior to the reference device is performed by a combination of careful circuit  analysis and empirical adjustment by signal measurements.
Although the filter-waveshaper approach is extendable to block-oriented gray-box modeling \cite{eichas2016black, eichas2017virtual, eichas2018gray}, this empirical process can be time-consuming and prone to errors.

Previous work on neural network models for guitar amplifiers has proposed high-quality real-time solutions 
\cite{wright2019-dafx-real-time-black-box-rnn, wright2019-real-time-gru-dafx, wright2020-guitar-amp-deep-learning-journal,schmitz2018-tube-amp-lstm}, but the research so far has limited itself on learning an input-output mapping at static control settings. Although there has been initial work on controllable neural amp modeling \cite{damskagg2019-icassp-deep-learning-tube-amp}, the results were limited to a single control variable and relied on a simulated ground truth to turn the control knob virtually.

While the present neural amp modeling techniques manage to capture single control settings, reproducing the full range of behaviour on varying amplifier controls presents further challenges for digital models.
A faithful model should recreate the behaviour of the amplifier at all combinations of the controls, including potentiometer response curves and complex interactions between various control settings. Moreover, the combinatorial space containing the various control settings grows exponentially with the number of controls, presenting a challenge not only for the modeling, but also for the perceptual validation of the model.
Manually verifying the model accuracy at a handful of control positions can only cover a tiny fraction of the control space in a reasonable amount of time.
Therefore, modeling physical amplifiers at their full range of controls using black-box methods requires an efficient and reliable data collection pipeline.

This paper outlines a data-driven approach to making controllable neural guitar amplifier models. In addition to the novel problem formulation, the paper describes an automated data-collection pipeline suitable for physical amplifier devices. The approach is validated by using the data to train a controllable neural network model and comparing the subjective quality to the reference and an offline white-box SPICE circuit model.

The paper is structured as follows.
Section \ref{sec:problem_formulation} defines the data driven guitar amplifier modeling problem in a concise form, and expands on the practical considerations of collecting data from real amplifiers. Section \ref{sec:model} briefly describes neural network architectures suitable for the guitar amp modeling task. Finally, Section \ref{sec:eval} outlines the experiments on training an evaluating a controllable neural network amp model.

\section{Data-driven amplifier models with controls}
\label{sec:problem_formulation}

\begin{figure}[tb]
\includegraphics[width=0.8\linewidth]{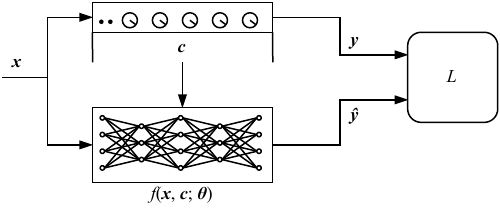}
\centering
\caption{Neural network amplifier model training scheme. A model $f$ 
receives control positions $\mathbf{c}$ as conditioning, and maps an input signal $\mathbf{x}$ to resemble a target output signal $\mathbf{y}$. A loss function $L$ scores the similarity of model output $\hat{\mathbf{y}}$ and $\mathbf{y}$ and provides a learning signal to adjust the model parameters to improve the resemblance.}
\label{fig:amp-nn}
\end{figure}

An overview of the proposed data-driven framework for amplifier modeling is illustrated in Figure~\ref{fig:amp-nn}.
The overall goal of modeling is to capture the full range of an amplifier's response to any given audio input, and enable continuous adjustment of the amplifier's controls, such as knobs and switches.
In practice, it is sufficient to sample the amplifier's control space at a discrete collection of control positions and train a neural network model that learns to generalize to the continuum of unseen control positions.
As such, a single training example in the dataset contains 
\begin{itemize}
    \item $\mathbf{x}^{(i)}$ -- an input audio segment, typically  guitar or bass playing
    \item $\mathbf{y}^{(i)}$ -- a recorded amplifier response to the input, used as a target for the model output 
    \item $\mathbf{c}^{(i)}$ -- control values describing the amplifier settings
\end{itemize}
These variables are grouped together as a triple $(\mathbf{x}^{(i)}, \mathbf{y}^{(i)}, \mathbf{c}^{(i)})$ to constitute the $i$-th training example in a dataset.
That is, the collection of $N$ training examples forms the dataset $\mathcal{D}$ as
\begin{equation}
    \mathcal{D} = \left\{\left( \mathbf{x}^{(1)}, \mathbf{y}^{(1)}, \mathbf{c}^{(1)} \right), \dots,
    \left( \mathbf{x}^{(N)}, \mathbf{y}^{(N)}, \mathbf{c}^{(N)} \right) \right\}_{i=1}^{N}.
\end{equation}
The control values can either be continuous, in the case of
representing a knob position, or discrete, in the case of
representing a switch position. Furthermore, the controls can be time-varying, but usually change at a much slower rate than the audio rate signals.

Our primary model of interest is a neural network that accepts $\mathbf{x}$ and $\mathbf{c}$ as input, and is parameterized by $\bm\theta$, i.e., $\hat{\bf{y}} = f(\mathbf{x}, \mathbf{c}; \bm{\theta})$.
Using a loss function $L(\hat{\mathbf{y}}, \mathbf{y})$ that measures the discrepancy between the model predictions $\hat{\mathbf{y}}$ and targets $\mathbf{y}$, we 
utilize standard supervised learning with stochastic gradient descent 
to learn the network parameters $\bm{{\theta}}$ that minimize the average loss over our training set
\begin{equation}
\bm{\theta}^* := \argminA_{\bm{\theta}} \frac{1}{N}\sum_{i=1}^{N}L(f(\mathbf{x}^{(i)}, \mathbf{c}^{(i)}; \bm{\theta}), \, \mathbf{y}^{(i)}).
\end{equation}
Given a sufficiently large and representative training set, and an appropriate model, the model generalizes to inputs outside the training set and learns to interpolate the behaviour of unseen control settings.

\subsection{Robotics for data collection}
\label{sec:data}

\begin{figure}[thb]
\includegraphics[width=1.0\linewidth]{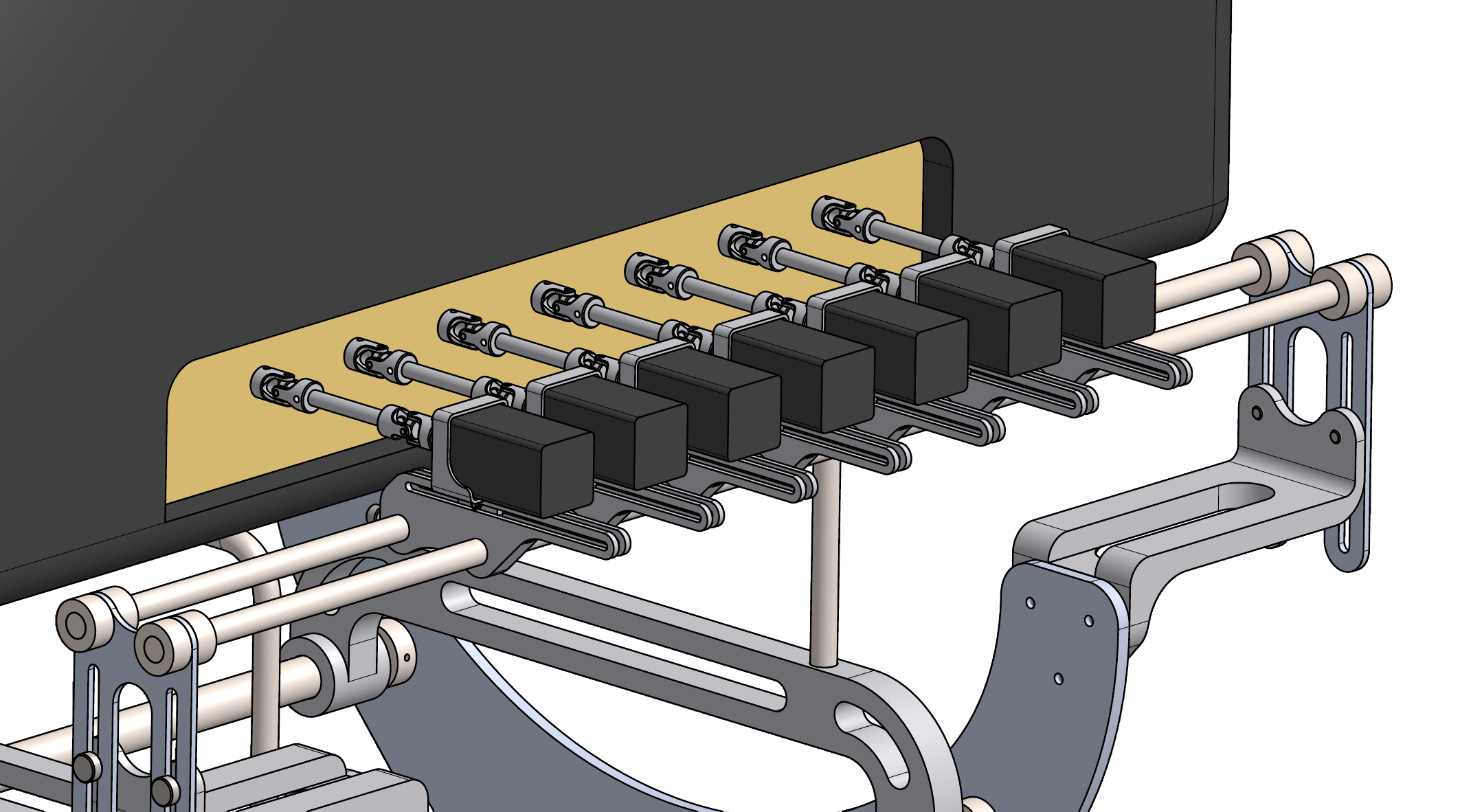}
\centering
\caption{Data collection robot connected to an amplifier~\cite{tina_patent_eu, tina_patent_us}.}
\label{fig:tina}
\end{figure}
Data-driven models depend on high quality data for the best results. In the case of amplifier modeling, adjusting and recording the control positions requires a degree of precision, consistency, and repetitiveness. Simply, this is not feasible for a human operator. In contrast, the task is well suited for a robot. As such, we have designed and built a robot for the very task of automated data collection from guitar amplifiers~\cite{tina_patent_eu, tina_patent_us}. 
 Figure \ref{fig:tina} illustrates the robot connected to an amplifier.

As seen in Figure \ref{fig:tina}, electric motors are attached to each relevant control of a physical amplifier.
By controlling the motors, the knobs on the amplifier can be set to any configuration.
Furthermore, the robot is connected to an audio interface, such that audio can be played and recorded while keeping track of the knob positions.
During data set collection, the robot moves the controls to different positions, plays audio through the amplifier and records the output.
As input audio material to the amplifier, we use a large collection of guitar, bass, and synthetic recordings, which are randomly sampled for each training segment.
To ensure good generalisation, it is important that the source material exhibits as much variation as possible, while being representative of the different types of signals that are expected to be played through the reference device or the model.

\subsection{Traversing the control space}
The amplifier control space is continuous, but any practical measurements are confined to a finite collection of control positions. 
Therefore, we need to first devise a sampling strategy to collect sufficient information about the controls in a finite number of samples.
Perhaps the simplest approach is to sample each knob at a number of discrete positions, and to go through all possible combinations.
However, this fixed grid approach very soon runs into issues with exponential scaling. 
For example, let's consider a case where each knob is discretized to 10 positions.
An amplifier with a single knob, say, a \emph{volume} control, would only require ten positions, while adding a \emph{gain} control would increase the number of combinations to 100. It is not uncommon for amplifiers to have six or more knobs, which would result in over a million combinations on a regular grid (in general $10^n$ recordings for $n$ knobs).
Furthermore, a switch with $m$ positions, increases the number of recordings by a factor of $m$.

Not only is the fixed grid approach rigid in terms of number of samples, but it runs a risk of overfitting to the grid points.
Instead, we can break the symmetries of the regular grid and freely choose the number of data points by applying a random sampling strategy. 
In a randomized sampling strategy, control positions are chosen from a uniform distribution for each control, resulting in an unbiased sampling of the overall control space across the whole data set.
The number of data points can be chosen freely to strike a balance between a sufficiently dense sampling of the control space and resource constraints, such as disk storage and recording time.

When sampling a physical device, we care not only about the density with which we sample the control space, but also how this process affects the wear and tear of the involved mechanical components, and the time it takes to move the knobs between different positions.
Thus, a strategy which minimizes the total travel required to sample all positions while balancing the travel per component is needed. 
To address this issue, we combine the random sampling procedure with an optimized sorting approach. 
In particular, all measured control configurations are generated ahead of time, and a sorting of this list is designed to minimize the overall distance travelled. 
By further assuming that we start and finish the recording with all controls at zero, finding the optimal path through random samples becomes a traveling salesman problem (TSP). 

Subject to our application, we first define an appropriate distance measure between the different control configurations, i.e., different control vectors $\mathbf{c}$. As we are concerned about the overall distance travelled by each component, a natural choice is the $\textrm{L}_1$ distance. 
More formally, given the desired number of examples $N$  we compute the matrix of control-position-wise distances $\mathbf{D} \in \mathbb{R}^{N \times N}$ as
\begin{equation}
    \mathrm{D}_{[i, j]} = ||\mathbf{c}^{(i)} - \mathbf{c}^{(j)}||^{}_{1}, \, \, \{\forall \, i, j \in \mathbb{N} \, | \, i, j \in [1, N] \} \, ,
\end{equation}
where $i$ and $j$ are indices of two sampled candidate vectors that contain the control positions.

\begin{figure}[thb]
     \centering
     \begin{subfigure}[t]{0.49\linewidth}
         \centering
         \includegraphics[width=\linewidth]{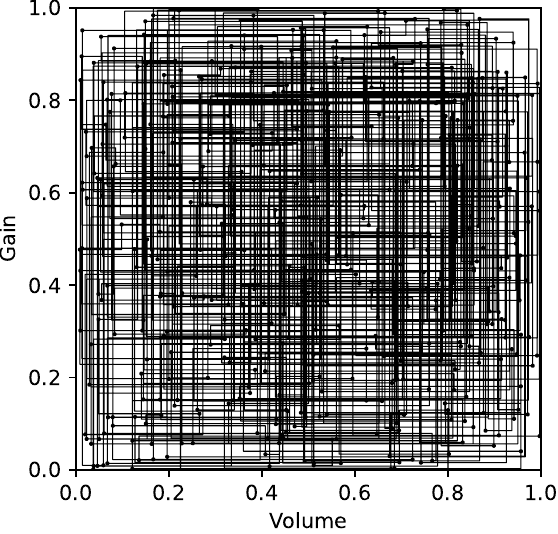}
         \caption{Random path}
         \label{fig:pathfinding-input}
     \end{subfigure} %
     \begin{subfigure}[t]{0.49\linewidth}
         \centering
         \includegraphics[width=\linewidth]{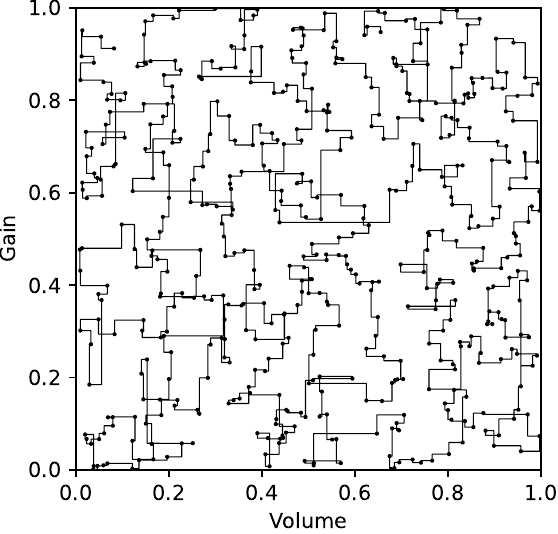}
         \caption{Sorted path}
         \label{fig:pathfinding-sorted}
     \end{subfigure}
     \caption{Example pathfinding solution for the case of two knobs and 500 data points. The data points are shown as dots, and the line segments between them show the travelled path for a) random order, and b) sorted order using a TSP solution.}
     \label{fig:shp}
\end{figure}

Although exact solutions of the traveling salesman problem are NP hard, we can instead use an efficient polynomial-time approximation \cite{christofides1976-worst-case-analysis-traveling-salesman}.
Figure \ref{fig:shp} shows an example pathfinding solution in the case of two knobs.

\section{Neural models for guitar amplifier circuits}
\label{sec:model}
For neural network based nonlinear circuit modeling, CNNs \cite{damskagg2019-icassp-deep-learning-tube-amp, damskagg2019-smc-real-time-modeling-distortion-circuits} and RNNs \cite{covert2013vacuum, zhang2018vacuum, schmitz2018nonlinear, wright2019-dafx-real-time-black-box-rnn, peussa2021exposure} have been applied successfully.
Furthermore, various architectures more specific to audio circuit modeling have been proposed \cite{ramirez2019-modeling-nonlinear-audio-effects-neural-nets-icassp, nercessian2021-differentiable-biquads, parker2019-dafx-state-transition-network}.

To draw an analogy to the familiar DSP models used in digital amplifier modeling, a CNN can be seen as a network of finite impulse response (FIR) filters and nonlinear activation functions.Convolution neural networks are essentially a trainable multichannel extension of the filter-waveshaper paradigm commonly used in audio signal processing.   
In particular, a causal convolution layer can be written in the time domain as
\begin{equation}
    \mathbf{y}_t = f\left(\sum_{i=0}^{n} \mathbf{W}_i \mathbf{x}_{t-i} + \mathbf{b}\right),
\end{equation}
where $\mathbf{x}_{t} \in \mathbb{R}^{q}$ is the input signal at time $t$, $\mathbf{W}_i \in \mathbb{R}^{p \times q}$ are the learned filter coefficients, $\mathbf{b} \in \mathbb{R}^{p}$ is a learned bias term, and $f(\cdot)$ is a nonlinear function applied element-wise.
In the case of amplifier modeling, the use of causal convolutions is motivated not only by the underlying causality of the device-under-test, but also by algorithmic latency concerns for the model itself.

To continue the analogy of convolution networks as extended feedforward FIR filter systems, RNNs can be seen as the neural counterpart to feedback IIR filter systems. 
Furthermore, RNNs can be interpreted as the discretisation of a system of nonlinear ODEs \cite{tallec2018can, peussa2021exposure}, which are commonly used to represent the behaviour of nonlinear electric circuits. In general, given an excitation signal $\mathbf{x}_t$ (e.g., clean guitar) and the model state $\mathbf{h}_{t-1}$, an RNN updates its state to $\mathbf{h}_t$ and derives an output variable $\mathbf{y}_t$ (e.g., amplifier response to input) from the state   
\begin{equation}
     \mathbf{h}_t  = f (\mathbf{x}_t, \mathbf{h}_{t-1}),\ \mathbf{y}_t = g(\mathbf{h}_t).
\end{equation}

The amplifier controls can be included as an additional input vector $\mathbf{c}_t$ to the neural network, allowing a single model to represent different user control configurations.

\subsection{Loss functions}
The loss function measures the difference between the model prediction and the target signal. A common starting point is the mean-squared error (MSE)
\begin{equation}
    L_{\textrm{MSE}}(\hat{\mathbf{y}}, \mathbf{y}) = \frac{1}{B T} \sum_{i=1}^{B} \sum_{t=1}^{T}
    || \mathbf{y}_{t}^{(i)} - f(\mathbf{x}_{t}^{(i)} ,  \mathbf{c}^{(i)}) ||_{2}^{2}
\end{equation}
where $T$ is the number of time steps in a training example, and $B$ is the number of elements in a minibatch. This loss function corresponds to minimising the energy of the model's error
\begin{equation}
    \mathbf{e}_{t}^{(i)} = \mathbf{y}_{t}^{(i)} - f\left(\mathbf{x}_{t}^{(i)}, \mathbf{c}^{(i)}\right).
\end{equation}
An extension to MSE is to normalize the loss by the minibatch target signal energy, which leads to the error-to-signal ratio (ESR) loss \cite{damskagg2019-icassp-deep-learning-tube-amp}
\begin{equation}
\label{eq:esr-loss}
    L_{\textrm{ESR}}(\hat{\mathbf{y}}, \mathbf{y}) = \frac{\sum_{i=1}^{B} \sum_{t=1}^{T}
    ||\mathbf{e}_{t}^{(i)}||_{2}^{2}}{\sum_{i=1}^{B} \sum_{t=1}^{T} ||\mathbf{y}_{t}^{(i)}||_{2}^2}.
\end{equation}

\section{Experiments}
\label{sec:eval}

\subsection{Dataset}

For this paper we chose to model the Matchless DC-30\texttrademark amplifier, which can be considered a modern boutique version of the classic Vox AC-30. The device is a point-to-point hand-wired tube-powered combo amplifier, and it has five continuous controls (\textit{volume}, \textit{bass}, \textit{treble}, \textit{tone cut}, and \textit{master volume}) which we need to recreate.

Training examples were recorded with our purpose-built data collection robot. 
Each example in the dataset consists of a pair of one second long input-target audio segments, sampled at 48\,kHz. The control positions are kept constant over each segment, and stored together with the audio, as described in section \ref{sec:data}.
Input audio sequences were drawn randomly from a collection of guitar and bass recordings. The dataset totals around 4.5 hours of paired audio, randomly split into 15000 training and 1000 validation examples.

\subsection{Neural network model}

To focus on the data pipeline aspect of this paper, we adapt a model already present in the literature to show the feasibility of the modeling approach. LSTM models have already demonstrated their worth  in static snapshot models for tube amplifiers \cite{wright2019-dafx-real-time-black-box-rnn, schmitz2018nonlinear}.

\begin{figure}[thb]
     \centering
     \begin{subfigure}[t]{0.49\linewidth}
         \centering
         \includegraphics[width=\linewidth, trim=10 10 5 5, clip]{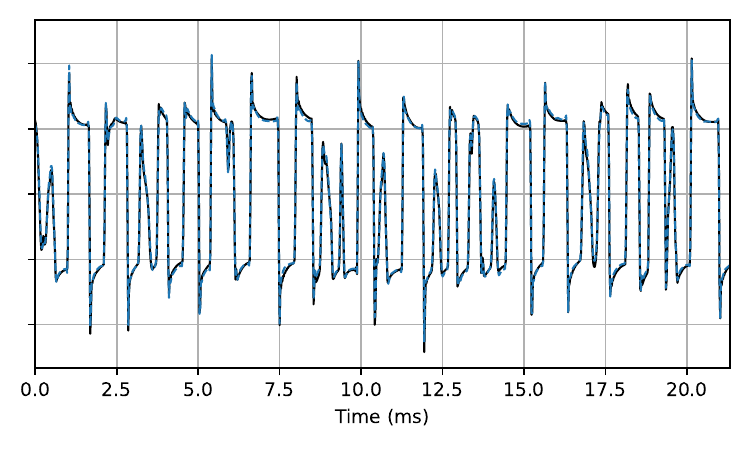}
         \label{fig:waveform-vox}
     \end{subfigure}
     \begin{subfigure}[t]{0.49\linewidth}
         \centering
         \includegraphics[width=\linewidth, trim=10 10 5 5, clip]{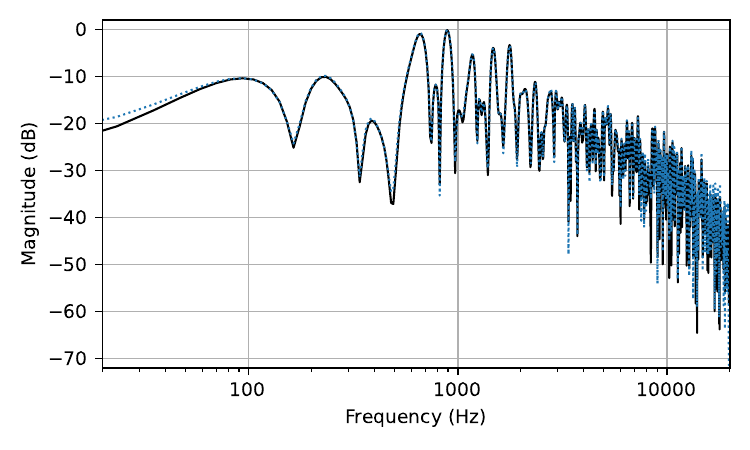}
         \label{fig:spectrum-vox}
     \end{subfigure}
     \caption{Comparison of model output (blue) to reference (black) in time  domain (left) and frequency domain (right).}
     \label{fig:vox}
\end{figure}

\begin{figure}[thb]
     \centering
     \begin{subfigure}[t]{0.49\linewidth}
         \centering
         \includegraphics[width=\linewidth,trim=10 10 5 5, clip]{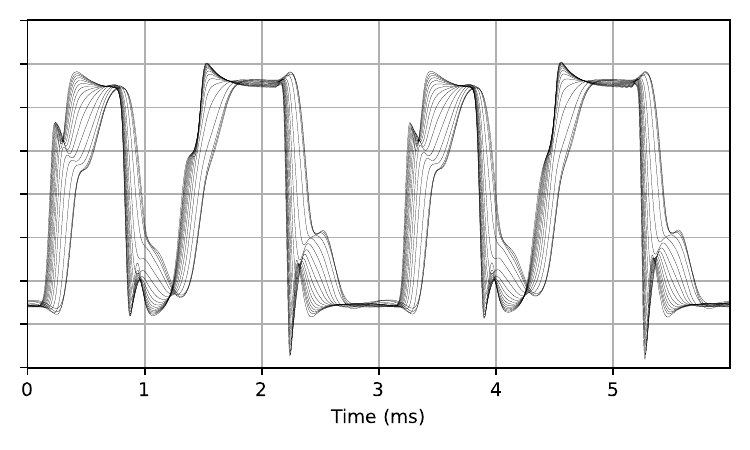}
         \label{fig:waveform-vox-controls}
     \end{subfigure}
     \begin{subfigure}[t]{0.49\linewidth}
         \centering
         \includegraphics[width=\linewidth,trim=10 10 5 5, clip]{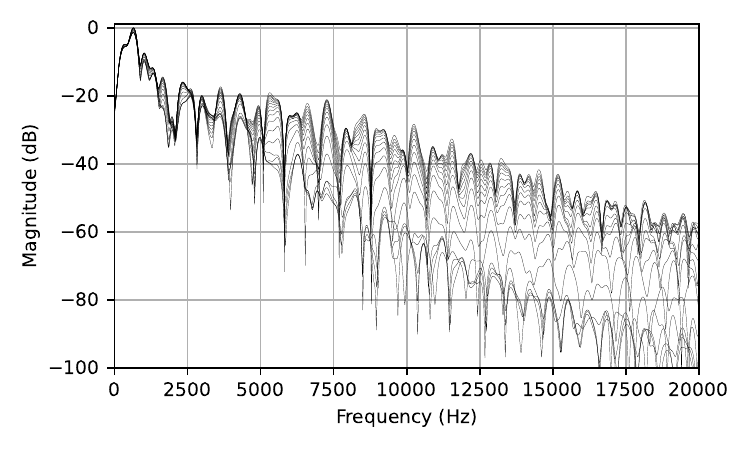}
         \label{fig:spectrum-vox-controls}
     \end{subfigure}
     \caption{Output from the \blackboxamp model with various \textit{tone cut} settings in time domain (left), and in frequency domain (right).}
     \label{fig:vox-controls}
\end{figure}

To incorporate the conditioning variables to the model, we normalize them to the range $[0, 1]$, and concatenate them as additional input channels to the LSTM network. Based on preliminary experiments, we found that a single layer LSTM with 32 cells (denoted LSTM-32) provides a good balance between perceptual quality and real-time cost. For the listening test, we trained the LSTM-32 model for 1M iterations 
using the ESR loss (Eq. \ref{eq:esr-loss}) and the Adam optimizer \cite{kingma2015-adam}. 
%
Figure \ref{fig:vox} shows a comparison between the model and the reference for an example guitar signal. Figure \ref{fig:vox-controls} shows the effect of the \textit{tone cut} control to the model response in time and frequency domain.

\subsection{SPICE model}


As a computationally optimistic white-box model, we recreated the amplifier schematic in LTSpice, which can be considered an industry standard tool in electrical circuit simulation.
This modeling approach requires a high amount of expert work, as it involves reverse engineering the circuit schematic and measuring each of the electrical component values on the target device.
In addition to the standard LTSpice components, our SPICE model used the 
Koren vacuum tube model \cite{koren1996-tube-model} with manual adjustment of the tube section input gains to match the saturation curves measured on the amplifier.

Test set audio samples were rendered with the LTSpice transient analysis tool using a variable step size, and resampled to a constant sample rate for listening. In terms of fidelity, the present SPICE model represents a best-case scenario for white box modeling, and is far from real time feasibility -- each of test samples took several hours to render.
While the amplifier under investigation is still amenable for offline processing, adding more components (such as tube gain stages for high gain amplifiers) quickly runs into issues with quadratic scaling in the circuit node connectivity graph.


\subsection{Listening test}

To evaluate the subjective quality of the amp models, we conducted a Difference Mean Opinion Score (DMOS) \cite{itu-rec1996-mos-dmos} listening test. In the test, a single test case presents the listener with a reference and a test sample, and asked the listener to rate how closely the test sample resembles the reference on a scale from 1 (bad) to 5 (excellent).
The test comprised five evaluation cases\footnote{
Readers can make their own judgements on the test material at { \url{https://neural-dsp-publications.github.io/demo-page-2022/}}. 
}, each rated by 30 expert listeners using headphones.
The test was conducted remotely using the WebMushra \cite{schoeffler2018webmushra} platform modified for DMOS.
Ratings were collected for the LSTM-32 model, the SPICE model, as well as a hidden reference and a low anchor,
which was produced by applying a 3.5 kHz low-pass filter on the reference samples.

In order to grade the model generalisation to various control positions, the reference amplifier knob positions were adjusted to fit the musical context for each evaluation case. The SPICE model was adjusted to the test configurations by measuring the potentiometer resistances to each instance. 
To ensure consistency, we applied the same guitar cabinet impulse response on all the systems, and matched the loudness of each sample using the LUFS \cite{itu-rec2015-lufs} metric.

\begin{figure}[thb]
    \centering
    \includegraphics[width=0.75\linewidth]{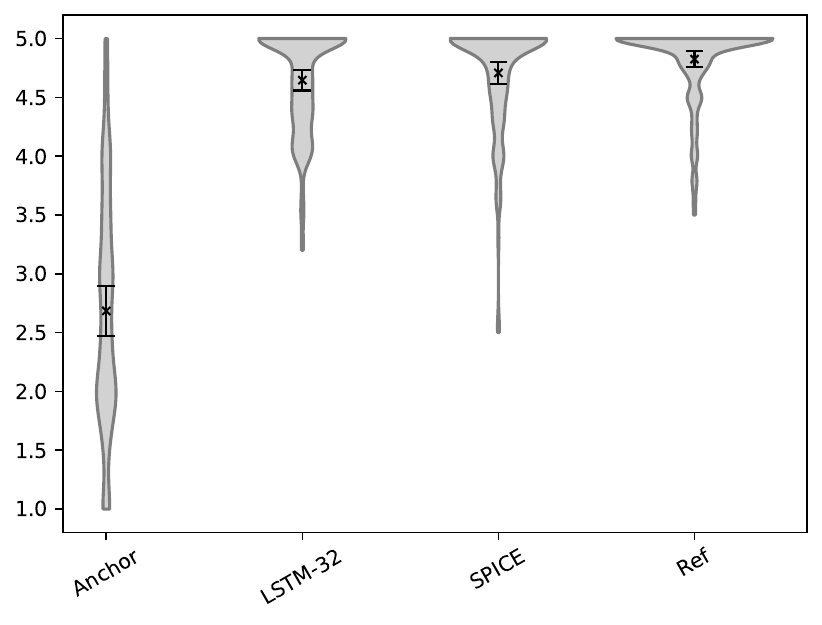}
    
    \caption{Listening test DMOS rating distributions, including mean ratings with 95\% confidence intervals based on the t-statistic. The results show high ratings for both an offline SPICE model, and a real-time LSTM neural network model trained using the proposed data driven framework. }
    \label{fig:listening-test-dmos}
\end{figure}

Fig. \ref{fig:listening-test-dmos} presents the listening test results. Both the SPICE model and the LSTM-32 neural network model achieve very high subjective quality.
A paired T-test with with Bonferroni correction for multiple comparisons indicated no statistically significant difference between the two models, while other pairings had significant differences.

\section{Conclusion}
This paper presents a data-driven approach to creating controllable neural network models of guitar amplifiers. 
While previous research on neural amp modeling has mostly constrained itself on static configuration snapshots, the present work extends the amp modeling problem to a full range of amplifier controls, discusses a practical implementation of data collection on physical devices, and derives a high quality neural network model from the data. Listening test results show that the resulting neural network model can match the subjective quality of an offline white-box SPICE circuit simulation.
\FloatBarrier


\begin{thebibliography}{10}

\bibitem{pakarinen2009-review-of-tube-amp-modeling}
J.~Pakarinen and D.~T. Yeh,
\newblock ``A review of digital techniques for modeling vacuum-tube guitar
  amplifiers,''
\newblock {\em Computer Music J.}, vol. 33, no. 2, pp. 85--100, 2009.

\bibitem{giampiccolo2021multiphysics}
R.~Giampiccolo, A.~Bernardini, G.~Gruosso, P.~Maffezzoni, and A.~Sarti,
\newblock ``Multiphysics modeling of audio circuits with nonlinear
  transformers,''
\newblock {\em Journal of the Audio Engineering Society}, vol. 69, no. 6, pp.
  374--388, June 2021.

\bibitem{najnudel2021power-balanced}
J.~Najnudel, T.~Hélie, D.~Roze, and R.~Müller,
\newblock ``Power-balanced modeling of nonlinear coils and transformers for
  audio circuits,''
\newblock {\em Journal of the Audio Engineering Society}, vol. 69, no. 7/8, pp.
  506--516, July 2021.

\bibitem{giampiccolo2022parallel}
R.~Giampiccolo, A.~Natoli, A.~Bernardini, and A.~Sarti,
\newblock ``Parallel wave digital filter implementations of audio circuits with
  multiple nonlinearities,''
\newblock {\em Journal of the Audio Engineering Society}, vol. 70, no. 6, pp.
  469--484, June 2022.

\bibitem{zoelzer2011dafx}
U.~Z{\"o}lzer, Ed.,
\newblock {\em DAFX: {Digital} {Audio} {Effects}},
\newblock Wiley, second edition, 2011.

\bibitem{eichas2016black}
F.~Eichas and U.~Z{\"o}lzer,
\newblock ``Black-box modeling of distortion circuits with block-oriented
  models,''
\newblock in {\em Proc. DAFx}, Brno, Czech Republic, 9 2016, pp. 39--45.

\bibitem{eichas2017virtual}
F.~Eichas, E.~Gerat, and U.~Z{\"o}lzer,
\newblock ``Virtual analog modeling of dynamic range compression systems,''
\newblock in {\em Proc. of the AES 142th Convention}, 5 2017.

\bibitem{eichas2018gray}
F.~Eichas and U.~Z{\"o}lzer,
\newblock ``Gray-box modeling of guitar amplifiers,''
\newblock {\em J. Audio Eng. Soc.}, vol. 66, no. 12, pp. 1006--1015, 12 2018.

\bibitem{wright2019-dafx-real-time-black-box-rnn}
A.~Wright, E.-P. Damsk{\"a}gg, V.~V{\"a}lim{\"a}ki, et~al.,
\newblock ``Real-time black-box modelling with recurrent neural networks,''
\newblock in {\em Int. Conference on Digital Audio Effects (DAFx)}, 2019.

\bibitem{wright2019-real-time-gru-dafx}
A.~Wright, E.-P. Damsk{\"a}gg, and V.~V{\"a}lim{\"a}ki,
\newblock ``Real-time black-box modelling with recurrent neural networks,''
\newblock in {\em Proc. DAFx}, Birmingham, UK, 9 2019, pp. 173--180.

\bibitem{wright2020-guitar-amp-deep-learning-journal}
A.~Wright, E.-P. Damsk{\"a}gg, L.~Juvela, and V.~V{\"a}lim{\"a}ki,
\newblock ``Real-time guitar amplifier emulation with deep learning,''
\newblock {\em Applied Sciences}, vol. 10, no. 3, 2 2020.

\bibitem{schmitz2018-tube-amp-lstm}
T.~Schmitz and J.-J. Embrechts,
\newblock ``Nonlinear real-time emulation of a tube amplifier with a long short
  time memory neural-network,''
\newblock in {\em Proc. AES 144th Convention}, 2018.

\bibitem{damskagg2019-icassp-deep-learning-tube-amp}
E.-P. Damsk{\"a}gg, L.~Juvela, E.~Thuillier, and V.~V{\"a}lim{\"a}ki,
\newblock ``Deep learning for tube amplifier emulation,''
\newblock in {\em IEEE Int. Conference on Acoustics, Speech and Signal
  Processing (ICASSP)}, Brighton, UK, 2019, pp. 471--475.

\bibitem{tina_patent_eu}
D.~A. Castro-Borquez, A.~T. Peussa, A.~Gotsopoulos, E.-P. Damsk{\"a}gg,
  L.~Juvela, K.~E.~A. Rauhanen, and T.~W. Sherson,
\newblock ``Robotic system for controlling audio systems,'' Feb. 2021,
\newblock EU Patent Application No. 1156974.4 - 1207.

\bibitem{tina_patent_us}
D.~A. Castro-Borquez, A.~T. Peussa, A.~Gotsopoulos, E.-P. Damsk{\"a}gg,
  L.~Juvela, K.~E.~A. Rauhanen, and T.~W. Sherson,
\newblock ``Robotic system for controlling audio systems,'' Feb. 2021,
\newblock US Patent Application No. 17/669,797.

\bibitem{christofides1976-worst-case-analysis-traveling-salesman}
N.~Christofides,
\newblock ``Worst-case analysis of a new heuristic for the travelling salesman
  problem,''
\newblock Tech. {R}ep., Carnegie-Mellon Univ Pittsburgh Pa Management Sciences
  Research Group, 1976.

\bibitem{damskagg2019-smc-real-time-modeling-distortion-circuits}
E.-P. Damsk{\"a}gg, L.~Juvela, V.~V{\"a}lim{\"a}ki, et~al.,
\newblock ``Real-time modeling of audio distortion circuits with deep
  learning,''
\newblock in {\em Proc. Sound and Music Computing Conference (SMC)}, Malaga,
  Spain, 2019, pp. 332--339.

\bibitem{covert2013vacuum}
J.~Covert and D.~L. Livingston,
\newblock ``A vacuum-tube guitar amplifier model using a recurrent neural
  network,''
\newblock in {\em Proc. IEEE Southeastcon}, 2013.

\bibitem{zhang2018vacuum}
Z.~Zhang, E.~Olbrych, J.~Bruchalski, T.~J. McCormick, and D.~L. Livingston,
\newblock ``A vacuum-tube guitar amplifier model using long/short-term memory
  networks,''
\newblock in {\em Proc. IEEE SoutheastCon}, 2018.

\bibitem{schmitz2018nonlinear}
T.~Schmitz and J.-J. Embrechts,
\newblock ``Nonlinear real-time emulation of a tube amplifier with a long short
  time memory neural-network,''
\newblock in {\em Audio Engineering Society Convention}, 2018.

\bibitem{peussa2021exposure}
A.~Peussa, E.-P. Damsk{\"a}gg, T.~Sherson, S.~Mimilakis, L.~Juvela,
  A.~Gotsopoulos, and V.~V{\"a}lim{\"a}ki,
\newblock ``Exposure bias and state matching in recurrent neural network
  virtual analog models,''
\newblock in {\em Proc. Int. Conference on Digital Audio Effects (DAFx)}, 2021,
  pp. 284--291.

\bibitem{ramirez2019-modeling-nonlinear-audio-effects-neural-nets-icassp}
M.~A.~M. Ram{\'\i}rez and J.~D. Reiss,
\newblock ``Modeling nonlinear audio effects with end-to-end deep neural
  networks,''
\newblock in {\em Int. Conference on Acoustics, Speech and Signal Processing
  (ICASSP)}, 2019, pp. 171--175.

\bibitem{nercessian2021-differentiable-biquads}
S.~Nercessian, A.~Sarroff, and K.~J. Werner,
\newblock ``Lightweight and interpretable neural modeling of an audio
  distortion effect using hyperconditioned differentiable biquads,''
\newblock in {\em Proc.~{ICASSP}}, 2021, pp. 890--894.

\bibitem{parker2019-dafx-state-transition-network}
J.~D. Parker, F.~Esqueda, and A.~Bergner,
\newblock ``Modelling of nonlinear state-space systems using a deep neural
  network,''
\newblock in {\em Proc.~{DAFx}}, 2019.

\bibitem{tallec2018can}
C.~Tallec and Y.~Ollivier,
\newblock ``Can recurrent neural networks warp time?,''
\newblock in {\em Proc. Int. Conference on Learning Representations (ICML)},
  Vancouver, Canada, 2018.

\bibitem{kingma2015-adam}
D.~P. Kingma and J.~Ba,
\newblock ``Adam: A method for stochastic optimization,''
\newblock in {\em Proc. ICLR}, 2015.

\bibitem{koren1996-tube-model}
N.~Koren,
\newblock ``Improved vacuum tube models for {SPICE} simulations,''
\newblock {\em Glass Audio}, vol. 8, no. 5, pp. 18--27, 1996.

\bibitem{itu-rec1996-mos-dmos}
I.-T. Rec,
\newblock ``{P.~800}: Methods for subjective determination of transmission
  quality,''
\newblock {\em International Telecommunication Union, Geneva}, vol. 22, 1996.

\bibitem{schoeffler2018webmushra}
M.~Schoeffler, S.~Bartoschek, F.-R. St{\"o}ter, M.~Roess, S.~Westphal,
  B.~Edler, and J.~Herre,
\newblock ``{WebMUSHRA} -- {A} comprehensive framework for web-based listening
  tests,''
\newblock {\em J. Open Research Software}, vol. 6, no. 1, 2 2018.

\bibitem{itu-rec2015-lufs}
I.-T. Rec,
\newblock ``{BS.~1770}: Algorithms to measure audio programme loudness and
  true-peak audio level,''
\newblock {\em International Telecommunication Union, Geneva}, 2015.

\end{thebibliography}

\end{document}